\documentclass[10pt,a4paper,margin=1in]{article}
\usepackage[paper=a4paper,margin=1in]{geometry}
\usepackage[T1]{fontenc}
\usepackage{amsfonts}
\usepackage{amsmath}
\usepackage[pdftex]{graphicx} 
\usepackage[pdftex,linkcolor=black,pdfborder={0 0 0}]{hyperref} 
\usepackage{calc} 
\usepackage{enumitem} 
\usepackage{amsthm}
\usepackage{bm}
\usepackage{xcolor}
\usepackage{natbib}
\usepackage{placeins}
\usepackage{comment}
\usepackage{algorithm2e}
\RestyleAlgo{ruled}
\usepackage{dsfont}
\usepackage{pdflscape}
\usepackage{pdfpages}
\usepackage{adjustbox}
 \usepackage{hyperref}
 
\usepackage{amsmath}
\usepackage{amssymb}
\usepackage{graphicx}
\usepackage[english]{babel}
\usepackage{xcolor} 
\usepackage{acronym}

\acrodef{baggls}[BaGGLS]{\emph{Bayesian Group Global Local Shrinkage}}

\DeclareMathOperator{\ND}{\mathcal{N}}

\DeclareMathOperator{\BerD}{Bern}

\DeclareMathOperator{\GaD}{\mathcal{G}}

\DeclareMathOperator{\IGD}{\mathcal{IG}}

\DeclareMathOperator{\diag}{diag}  

\newcommand\KL[2]{\mathcal{D}_\text{KL}\left\lbrack #1 \,\vert\vert\, #2\right\rbrack}
\begin{document} 

\begin{titlepage}

\title{BaGGLS: A Bayesian Shrinkage Framework for Interpretable Modeling of Interactions in High-Dimensional Biological Data}

\author{Marta S. Lemanczyk$\mbox{}^{1\ast}$, Lucas Kock$\mbox{}^{2\ast}$, Johanna Schlimme$\mbox{}^1$,\\ Nadja Klein$\mbox{}^3$, and Bernhard Y. Renard$\mbox{}^1$} 

\date{\today}
\maketitle
\thispagestyle{empty}
\noindent

\begin{center}
{\Large Abstract} 
\end{center}
\vspace{-1pt}
\noindent Biological data sets are often high-dimensional, noisy, and governed by complex interactions among sparse signals. This poses major challenges for interpretability and reliable feature selection. Tasks such as identifying motif interactions in genomics exemplify these difficulties, as only a small subset of biologically relevant features (e.g., motifs) are typically active, and their effects are often non-linear and context-dependent. While statistical approaches often result in more interpretable models, deep learning models have proven effective in modeling complex interactions and prediction accuracy, yet their black-box nature limits interpretability.

We introduce BaGGLS, a flexible and interpretable probabilistic binary regression model designed for high-dimensional biological inference involving feature interactions. BaGGLS incorporates a Bayesian group global-local shrinkage prior, aligned with the group structure introduced by interaction terms. This prior encourages sparsity while retaining interpretability, helping to isolate meaningful signals and suppress noise. To enable scalable inference, we employ a partially factorized variational approximation that captures posterior skewness and supports efficient learning even in large feature spaces.

In extensive simulations, we compare BaGGLS to frequentist probit regressions (unconstrained and with L1-penalty) as well as a probit model with Markov Chain Monte Carlo (MCMC) sampling under a horseshoe prior. We can show that BaGGLS outperforms the other methods with regard to interaction detection and is many times faster than MCMC sampling under the horseshoe prior. We also demonstrate the usefulness of BaGGLS in the context of interaction discovery from motif scanner outputs (e.g., Find Individual Motif Occurrences (FIMO)) and noisy attribution scores from deep learning models. This shows that BaGGLS is a promising approach for uncovering biologically relevant interaction patterns, with potential applicability across a range of high-dimensional tasks in computational biology.

\vspace{20pt}
 

\vspace*{\fill}
\noindent {\small\textbf{Acknowledgments:} We gratefully acknowledge funding by the German Research Foundation (DFG) via the Research Unit KI-FOR 5363 (grant 459422098).}

\vspace{20pt}

\noindent{\small
$^1$ Hasso Plattner Institute, Digital Engineering Faculty, University of Potsdam, Potsdam, Germany\\
$^2$ Department of Statistics and Data Science, National University of Singapore\\
$^3$ Scientific Computing Center, Karlsruhe Institute of Technology, Karlsruhe, Germany\\
$^\ast$ These authors contributed equally; \href{mailto:marta.lemanczyk@hpi.de}{marta.lemanczyk@hpi.de}, \href{mailto:lucas.kock@nus.edu.sg}{lucas.kock@nus.edu.sg}
}

\end{titlepage}

\section{Introduction}
One of the greatest methodological challenges in the biomedical data domain is feature selection in high-dimensional data \citep{borah2024review, yang2021feature}. This challenge has historically been central in genomics, but with deep learning (DL) becoming the dominant analytical paradigm and post hoc attribution being routinely used for interpretability, the problem manifests in a different and more intricate form today. DL models excel at learning complex patterns in genomic sequences \citep{ismail2025deep}, However, the resulting models are highly nonlinear and operate in settings where biological features are both strongly interactive \citep{xie2025dna, forsberg2017accounting} and extremely sparse \citep{wheeler2016survey}. These properties amplify the already difficult task of interpreting which features matter and, crucially, how they interact.

As a result, we continue to struggle to reliably explain feature interactions \citep{borah2024review}. This difficulty is not merely algorithmic, but rooted in statistical challenges. Among these challenges are the curse of dimensionality, the tendency of interactions to explode combinatorically, sparsity of true effects, and high noise levels \citep{giraud2021introduction}. These challenges are in particularly pronounced when working with post hoc attribution maps. Attribution scores are often noisy and spurious \citep{majdandzic2023correcting}, and aggregating them to motif-level features produces high-dimensional, sparsely informative predictors whose reliability and error are difficult to determine. Altogether, this creates a setting in which classical feature selection approaches fall short, and even modern statistical approaches do not resolve the core problem reliably.

Consequently, we see a strong need for new methodology that addresses the specific structure of post hoc attribution in genomic deep learning. Here, we propose a structured binary regression model that incorporates a large number of potential main effects and interaction terms into its linear predictor. For interpretability and reliable identification of relevant effects, we impose sparsity on the coefficients. Within the Bayesian framework, sparsity is naturally enforced through shrinkage priors \citep{GeoMcc1997,IshRao2005,LiaPauMolClyBer2008,YanBonRei2025}. Many continuous shrinkage priors such as the Bayesian Lasso \citep{ParCas2008} and the horseshoe \citep{CarPolSco2010} follow a global–local structure \citep{PolSco2011}, in which global parameters jointly shrink coefficients toward zero while local parameters allow coefficient-specific deviations. 
\citet{XuSchMakQiaHop2017} extend this framework to include group based shrinkage parameters, facilitating structured regularization when coefficients can be meaningfully organized into groups. The inclusion of interaction terms results in many overlapping groups, where each group is formed of all terms involving a specific feature. To accommodate this structure, we extend the group-based shrinkage framework of \citet{XuSchMakQiaHop2017}, adapting it to handle the extremely large number of potential interaction terms present in genomic data. A detailed description of the resulting prior structure is provided in Section~\ref{subsec:prior}.

Exact Bayesian inference in the resulting high-dimensional Bayesian probit model is challenging and we propose a computationally efficient variational inference \citep[VI;~][]{BleKucMca2017} algorithm, a commonly applied technique in high-dimensional binary regression models \citep{ZhaXuZha2019,RaySzaCla2020}. Here, we extend the approach of \citet{FasDurZan2022} to our novel prior and consider a unified skew–normal \citep{AreAzz2006} approximation for the regression coefficients. This allows more flexibility than the commonly employed mean field approximation \citep[e.g.,~][]{DurRig2019}. The unified skew-normal distribution generalizes the Gaussian distribution to include skewness. Recently, skewness perturbed variational approximations were considered by several authors \citep[e.g.,~][]{TanChe2025,KocTanBanNot2025,PozDurSza2024} theoretically justified by the skewed Bernstein-von Mises theorem \citep{DurPozSza2024}. The unified skew-normal distribution is also a conjugate prior to the probit model \citep{Dur2019,AncFasDurZan2023}, so that the variational approximation can be efficiently learned using an analytic coordinate ascent updating scheme \citep[e.g.,~][]{Bis2006,RaySza2022}. This enables scalable inference even when the number of interactions is large.

In this paper, we introduce our novel method called \ac{baggls} in detail and demonstrate through extensive simulations that it outperforms state-of-the-art methods in both computational efficiency and interpretability. Importantly, we also show empirically that \ac{baggls} fills a methodological gap in post hoc analysis of genomic deep learning models. Many post hoc approaches attempt to understand learned regulatory mechanisms by detecting motifs through attribution scores \citep{novakovsky2023obtaining, van2024designing, bartoszewicz2021}. Attribution methods assign position-wise scores to sequences, and downstream tools such as TF-MoDISco \citep{shrikumar2018technical} extract motifs from these maps. However, these workflows do not capture interactions between motif patterns, even though such interactions are central to many genomic mechanisms \citep{xie2025dna}.

Motif scanners such as FIMO \citep{grant2011fimo} overcome some limitations by identifying matches to known PWMs from databases like JASPAR \citep{rauluseviciute2024jaspar}. Yet this produces a high-dimensional feature space containing hundreds of motifs, many of which produce spurious matches. When combined with noisy attribution scores, only a small subset of motifs, and often their interaction, contribute meaningfully to phenotypes. These conditions create exactly the type of high-dimensional, sparse, interaction-rich scenario where classical methods struggle. We show that \ac{baggls} successfully extracts these interacting signals and provides reliable interpretability.

The remainder of this paper is organized as follows. Section~\ref{sec:baggls} introduces \ac{baggls} and an efficient VI approach to posterior estimation. Section~\ref{sec:simulations} shows the merits of our approach empirically benchmarking it against state-of-the-art methods while Section~\ref{sec:application} illustrates its applicability to high-dimensional biological data along an application in interaction detection for genomic attribution scores. Finally, Section~\ref{sec:discussion} concludes. Code is available at \href{www.gitlab.com/dacs-hpi/baggls}{gitlab.com/dacs-hpi/baggls}.

\begin{figure}[tbh!]
    \centering
        \includegraphics[width=0.9\textwidth]{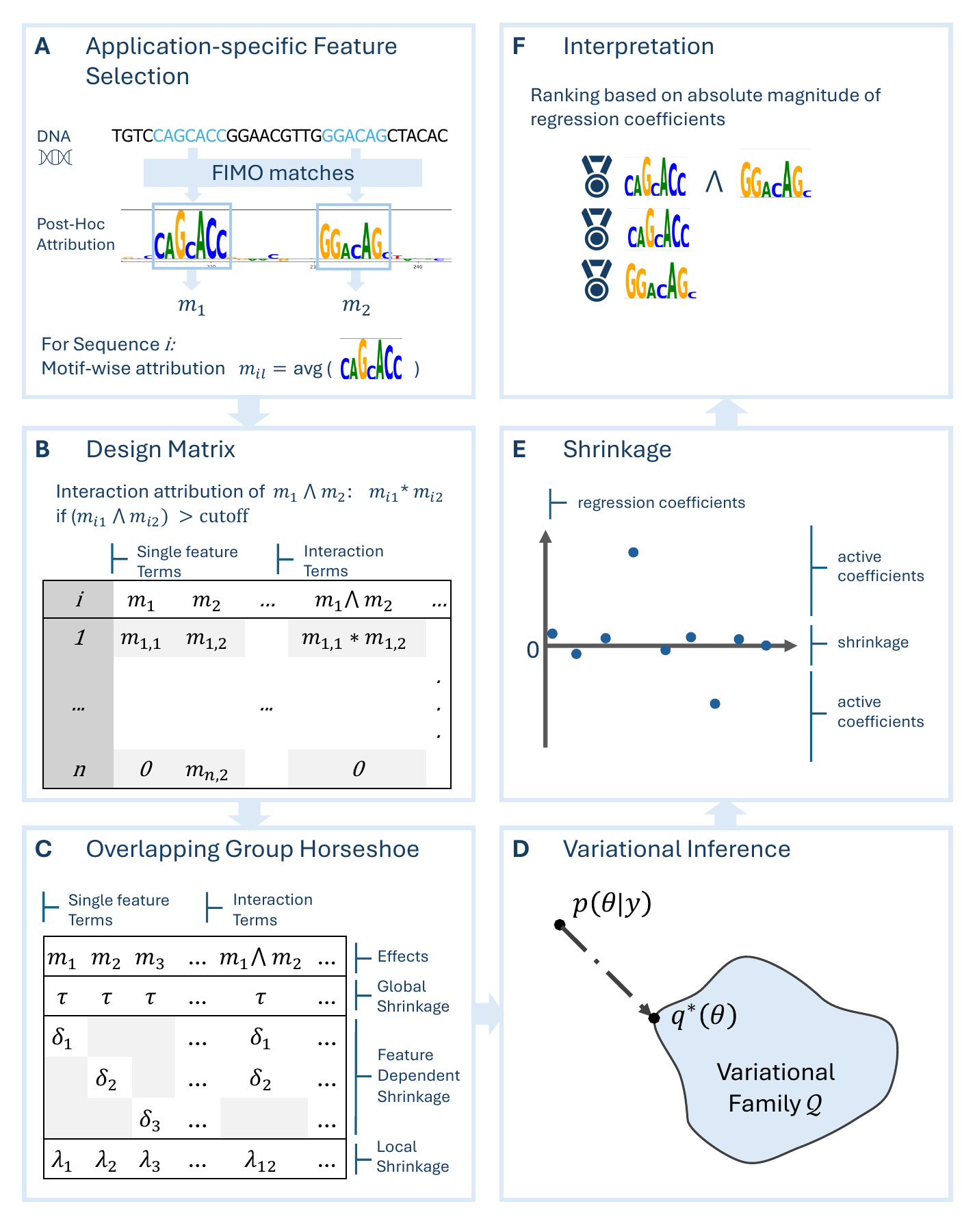}
    \caption{Schematic description of the \ac{baggls}-workflow. \textbf{A)} Feature selection based on the observed data. In our application discussed in Section~\ref{sec:application}, we use attribution scores from CNNs trained on FIMO matches. \textbf{B)} Based on expert knowledge a pre-defined set of potential effects including interpretable interactions is defined. \textbf{C)} Our novel overlapping group horseshoe prior matches the structure of the pre-defined effect terms and allows for global, local, and feature dependent shrinkage of the regression coefficients. \textbf{D)} Scalable Bayesian inference is carried our by projecting the true posterior onto a tractable family of variational posteriors $\mathcal{Q}$. \textbf{E)} The prior shrinks most regression coefficients towards $0$ resulting in a sparse and interpretable regression model. \textbf{F)} For interpretation, we propose to rank the effects based on absolute magnitude of their corresponding coefficients in the logistic regression model.
    }
    \label{fig:main}
\end{figure}

\section{Bayesian group global-local shrinkage} \label{sec:baggls}
We formally introduce the model in Section~\ref{subsec:model}. Our novel prior is introduced in Section~\ref{subsec:prior}, and Section~\ref{subsec:vb} describes our approach for scalable inference.

\subsection{Model formulation}\label{subsec:model}

We consider the probit model 
\begin{align} \label{eq: probit}
    y_i\mid\beta \sim \BerD(\Phi(x_i^\top\beta)) \qquad i=1,\dots n
\end{align}
where $\BerD(\pi)$ denotes the Bernoulli distribution with success probability $\pi$, $\Phi$ is the cumulative distribution function of the standard normal distribution, $x_i=(x_{i1},\dots,x_{ip})^\top$ is a vector of pre-defined effects including an intercept and interactions, and $\beta\in\mathbb{R}^p$ is a vector of model parameters to be learned. By introduction of latent variables $z=\left(z_1,\dots,z_n\right)^\top$ we can augment \eqref{eq: probit} as
\begin{align*}
    y_i = \mathcal{I}(z_i>0) , \qquad z_i\mid\beta \sim \ND(x_i^\top\beta,1),
\end{align*}
where $\ND(\mu,\sigma^2)$ denotes a normal distribution with mean $\mu$ and variance $\sigma^2$, and $\mathcal{I}(z_i>0)$ denotes the indicator function that takes a value of 1 if $z_i>0$ and 0 otherwise. This representation yields closed-form full-conditionals for $\beta$ and $z$, which will become useful for the computational efficient inference algorithm introduced later.
The linear predictor $x_i$ is derived from a $d$-dimensional vector $m_i=(m_{i1},\dots,m_{id})^\top$ of observed features, $i=1,\dots,n$, and contains not only linear effects, but also interaction terms of the $d$ features as illustrated in Figure~\ref{fig:main}B. For example, when considering an intercept, linear effects as well as all possible pairwise multiplicative interactions of the form $m_{il}m_{il'}$, $l\not=l'$, the linear predictor $x_i^\top\beta$ in \eqref{eq: probit} can be written as $\beta_0 + \sum_{l=1}^d \beta_l m_l + \sum_{l=1}^d\sum_{l'>l} \beta_{ll'}m_lm_{l'}$. Hence, $p$ is  typically much larger than $d$.  This is the structure we consider in Section~\ref{sec:application} when we apply \ac{baggls} to interaction discovery from motif scanner outputs. In this case $m_i$ will be a vector of attribution scores derived from a deep learning architecture. However, our general framework allows us to specify arbitrary effect and interaction terms. In this context, it is important to note that first, even for small and moderate $d$, the number of potential interactions is large and thus $x_i$ can be  high-dimensional. Second, we also  explicitly allow for much larger $p$ than the sample size $n$, that is, $p\gg n$. 

\subsection{Overlapping group horseshoe prior}\label{subsec:prior}
If the number of potential effect terms $p$ is large, \eqref{eq: probit} can be challenging to interpret. Based on common scenarios in biological applications, we make the following assumptions. (i) We assume that $\beta$ is  sparse. That is, most entries are zero and thus only a small number of effect terms, $x_{ij}$, influence $y_i$. (ii) We further assume, that only a small subset of features $m_{il}$ significantly influences $y$ and thus for most features all terms involving that feature are jointly zero.

The first assumption can be incorporated by considering an appropriate shrinkage prior. The prior structure also acts as an important regularization to the high-dimensional regression model. Here, we build on the popular horseshoe prior \citep{CarPolSco2010}. The horseshoe prior introduces a global shrinkage parameter $\tau$ controlling the overall level of shrinkage jointly for all predictors and local shrinkage parameters $\lambda_j$ $j=1,\dots,p$ controlling the shrinkage for the coefficient $\beta_j$ corresponding to effect $x_{ij}$. \citet{XuSchMakQiaHop2017} extend this prior structure to account for group-based shrinkage. The heredity assumption (ii) implies a structure with many overlapping groups. This would necessitate a prohibitively large number of hierarchies in the group horseshoe prior by \citet{XuSchMakQiaHop2017}. We thus propose the following alternation to their grouped global-local shrinkage prior that circumvents the need for many hierarchies for each predictor.

Let $J\in\mathbb{R}^{p\times d}$ be an indicator matrix where entry $J_{jl}$ indicates if feature $m_l$ contributes to term $x_j$. We propose the hierarchical prior
\begin{align*}
    \beta_j\mid \tau, \lambda,\delta &\sim\ ND\left(0,\tau \lambda_j \prod_{l=1}^d J_{jl}\delta_l\right),&\quad&j=1,\dots,p\\
    \tau\mid\nu&\sim\IGD\left(\frac{1}{2},\frac{1}{\nu}\right),\quad\nu\sim\IGD\left(\frac{1}{2},1\right),\\
    \lambda_j\mid c_j&\sim\IGD\left(\frac{1}{2},\frac{1}{c_j}\right),\quad c_j\sim\IGD\left(\frac{1}{2},1\right),&\qquad& j=1,\dots,p\\
    \delta_l\mid t_l&\sim\IGD\left(\frac{1}{2},\frac{1}{t_l}\right),\quad t_l\sim\IGD\left(\frac{1}{2},1\right),&\qquad& l=1,\dots,d,
\end{align*}
where $\delta=\left(\delta_1,\dots,\delta_d\right)^\top$, $\lambda=\left(\lambda_1,\dots,\lambda_p\right)^\top$. Here, $\IGD(\alpha,\beta)$ denotes an inverse Gamma distribution with shape parameter $\alpha$ and scale parameter $\beta$. As for the standard horseshoe prior $\tau$ controls global shrinkage, and $\lambda_j$ controls local shrinkage. In addition, $\delta_l$ controls joint group shrinkage for all effects including feature $m_l$. The structure of the indicator matrix $J$ informing the grouping structure depends on the effects considered in $x$ and needs to be specified upfront. To allow for consistent interpretation of the effect strength as well as consistent shrinkage through the shared shrinkage parameters we assume that all effects $x_{ij}, j=1,\dots,p$, are standardized to have unit scale across observations, $i=1,\dots,n$. A schematic description of our prior is given in Figure~\ref{fig:main}C. Each term has an individual local shrinkage parameter $\lambda_j$. The motif dependent group shrinkage parameter $\delta_l$ is active exactly for the terms involving the corresponding feature $m_l$. The parameter $\tau$ controls global shrinkage and is shared across all terms. The full vector of the $n+3p+2d+2$ unknown model parameters is $\theta=\left(z^\top,\beta^\top,\tau,\nu,\lambda^\top,c^\top,\delta^\top,t^\top\right)^\top$.

\subsection{Inference for large data sets}\label{subsec:vb}
Exact Bayesian inference in the high-dimensional probit model can be computationally challenging. VI emerged as a powerful alternative. The main idea illustrated in Figure~\ref{fig:main}D is to learn an approximation to the posterior density $p(\theta\mid y)$ using an approximating family of densities $\mathcal{Q}$. Most commonly, the optimal approximation $q^*(\theta)$ is chosen so that it minimizes the reverse Kullback-Leibler divergence, $$\KL{q(\theta)}{p(\theta\mid y))}=\mathbb{E}_{q}\left[\log q(\theta) - \log p(\theta\mid y)\right],$$where $\mathbb{E}_{q}[\cdot]$ denotes expectation with respect to $q(\theta)$, among all approximating families in $\mathcal{Q}$. A popular choice in logistic and probit regression models is the mean field assumption \citep[e.g.,~][]{DurRig2019}, which assumes independence between specific  blocks of $\theta$. This choice is computational efficient, but as recently shown by \citet{FasDurZan2022} can be too restrictive for large $p$. Therefore, the authors propose to relax the independence assumption between $\beta$ and $z$. This allows for more flexibility in the posterior approximation, as it yields a unified skew–normal \citep{AreAzz2006} approximation for the vector of regression coefficients. \citet{FasDurZan2022} consider a simple Gaussian prior for $\beta$. We extend their approach to the Gaussian scale mixture representation of the \ac{baggls} prior and consider variational approximations of the form 

\begin{align*}
    \mathcal{Q}=\Bigg\{q(\theta): q(\theta)=&q(\beta\mid z)\left(\prod_{i=1}^nq(z_i)\right)q(\tau)q(\nu)\left(\prod_{j=1}^p q(\lambda_j)q(c_j)\right)\left(\prod_{l=1}^d q(\delta_l)q(t_l)\right)\Bigg\}.
\end{align*}

We describe how $q^*(\theta)=\arg\min_{q(\theta)\in\mathcal{Q}}\KL{q(\theta)}{p(\theta\mid y))}$ can be derived in a computational attractive manner using a simple coordinate ascent algorithm. The final algorithm updates one variational factor at a time while holding the others fixed, cycling through coordinates until convergence. Under the variational family $\mathcal{Q}$ given above all updates are in closed-form (see Appendix~A).
It is possible to efficiently sample from $q(\beta)=\int q(\beta\mid z)q(z)dz$ even when $p\gg n$ using the strategies outlined in \citet{BhaChaMal2016} and the posterior mean $\widehat{\beta}=\mathbb{E}_{q}[\beta]$ is given in closed form. Due to the variational approximation, samples from $q^\ast(\theta)$ will not yield exact uncertainty quantification, and we use the point estimator $\widehat{\beta}$ for out of sample prediction and interpretation of the regression model. 

\section{Simulations} \label{sec:simulations}
\subsection{Illustrative example} \label{subsec:simulations1}
\paragraph{Data generating process} We generate $m=500$ data sets with $n=500$ independent observations. For $m_i=(m_{i1},\dots,m_{id})$ with $m_{ij}\sim \GaD(1,1)$, similar as in our application presented in Section~\ref{sec:application} the corresponding vector of predictors $x_i$ consists of an intercept, linear effects $m_j$, $j=1,\dots,d$, and all possible pairwise interactions $m_jm_{j’}$ for $j<j’$. We set $d=10$, so that $p=56$ in our first set-up. After standardization of the design matrix, we generate observations $y_i\sim\BerD(\Phi(x_i^\top\beta^\ast))$, where all entries of $\beta^\ast$ are zero except for the entries corresponding to $m_1$, $m_2$, and the interaction $m_1m_2$. The true regression coefficient $\beta^\ast$ is thus extremely sparse and  reflects the sparsity assumption described in Section~\ref{subsec:prior}.

\paragraph{Benchmark methods} 
We consider different regularization techniques for the probit model \eqref{eq: probit} that result in sparse and interpretable models. We use the same design matrix for all benchmarks, so that the models are directly comparable. In particular, we compare \ac{baggls} with the following benchmarks:
\begin{itemize}
    \item[] UC: Unconstrained frequentist probit regression fitted via maximum likelihood,
    \item[] L1: Frequentist probit regression with $L_1$-penalty and default hyperparameters as implemented in \texttt{statsmodels} \citep{SeaPer2010},
    \item[] HS: MCMC sampling for the probit model equipped with the horseshoe prior \citep{CarPolSco2010} as implemented in \texttt{brms} \citep{Bue2017}.
\end{itemize}

\begin{figure}[tbh!]
    \centering
    \includegraphics[width=0.9\linewidth,keepaspectratio]{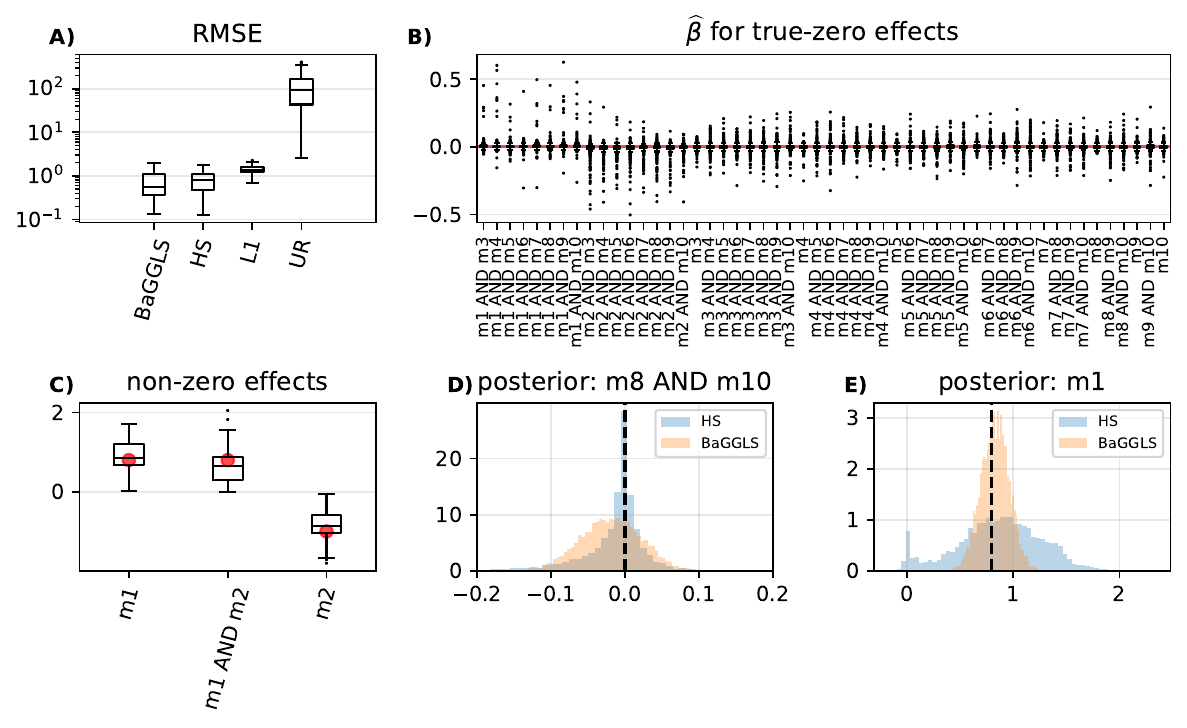}
    \caption{Simulations $n=500, d=10$. \textbf{A)} RMSE across 500 independent runs on log-scale. Lower values are prefered. Our propossed method outperforms the benchmarks in recovery of the true coefficient vectos \textbf{B)}~Boxplots for $\widehat{\beta}_j$ for all true-zero coefficients across 500 independent runs. For most repetitions, \ac{baggls} shrinks effects that do not contribute to the response successfully towards zero resulting in an interpretable model. \textbf{C)} Boxplots for $\widehat{\beta}_j$ for all non-zero coefficients. True values are marked by dots. \ac{baggls} correctly estimates the effect size for the three active effects in the data generating process. \textbf{D)} Samples from the estimated marginal posterior for one true-zero coefficient for HS and \ac{baggls}. Both posteriors are correctly centered around zero and skewed indicating that the skewed variational approximation used for \ac{baggls} is helpful in tightly approximating the true posterior. \textbf{E)} Samples from the estimated marginal posterior for one non-zero effect for HS and \ac{baggls}. Again both posteriors are centered around the true value. The variational approximation for \ac{baggls} is sharper than the posterior under HS.}
    \label{fig:sim_small}
\end{figure}

\paragraph{Results}
Figure~\ref{fig:sim_small}A shows boxplots of the root mean squared error (RMSE), $\text{RMSE}(\widehat{\beta})=\left(\sum_{j=1}^p(\beta^\ast_j-\widehat{\beta}_j)^2\right)^{1/2}$, over the 500 independent repetitions. \ac{baggls} has the smallest average RMSE (0.7022), followed by HS (0.8052), and L1 (1.3827). Figure~\ref{fig:sim_small}B plots estimates $\widehat{\beta}_j$ for all coefficients with $\beta^\ast_j=0$ showing effective shrinkage toward zero. As expected, the largest variability occurs for joint effects involving $m_1$ or $m_2$. Non-zero effects ($\beta^\ast_j\not=0$) are are accurately recovered (Figure~\ref{fig:sim_small}C). Figure~\ref{fig:sim_small}D shows a histogram for a representative non-informative coeffiencent under \ac{baggls} and under HS from their respective marginal posteriors. Both are centered near zero and on a similar scale. Notably, the marginal posterior $q(\beta_j)$ under \ac{baggls} is skewed, which is captured due to the unified skew-normal variational family. However, this family cannot capture the characteristic spike around zero. For a representative non-zero coefficient (Figure~\ref{fig:sim_small}E), \ac{baggls} concentrates more tightly around the truth, while HS posteriors are more dispersed. Nevertheless, both methods yield similar posterior means. Due to the variational approximation, posterior samples from \ac{baggls} will not yield exact uncertainty quantification. On this data, \ac{baggls} takes on average only $0.59$ seconds and is therefore more then 100 times faster then MCMC sampling under the horseshoe prior rendering it suitable for large scale applications.

\subsection{Scalability}

\paragraph{Data generating process} To further investigate the performance of \ac{baggls} we now vary the number of observations $n$ and the number of observed features $d$ in the data generative process described in Section~\ref{subsec:simulations1}. Due to the non-linear relationship between $d$ and the number of effects $p$ a slight increase in $d$ leads to a large increase in $p$ and thus to a much more challenging inference task. Here, we consider $n=500$ and $n=2,000$ as well as $d=10, 15, 20$ resulting in a total of $6$ scenarios. These values are chosen to reflect common scenarios in real world data and result in a small sample-to-feature ratio $n/p$. This matches the general structure of the genomic data considered in Section~\ref{sec:application}. 

\paragraph{Performance metrics} 
In addition to the overall RMSE, we consider several other performance metrics. First, we evaluate discrimination via the area under the receiver operating characteristic curve (AUC) and probabilistic accuracy via the Brier score, $n^{-1}\sum_{i=1}^n(\Phi(x_i^\top\widehat{\beta})-y_i)^2$, both evaluated on an additional hold-out test data set with $n=10,000$ observations. The Brier score is a proper scoring rule \citep{GneRaf2007}. These metrics are useful in comparing the out-of-sample predictive power of the different methods. Although prediction is not the primary objective of \ac{baggls}, reasonable out-of-sample performance remains desirable. However, when prediction is the main goal rather than interpretability, methods with weaker structural assumptions may be more suitable. To quantify effect recovery in high dimensions, we report RMSE computed separately over the active (nonzero) and inactive (zero) entries of $\beta^\ast$. In addition, we report the proportion of runs in which the three truly active effects are ranked among the top 20 and, separately, among the top 3 estimated effects by absolute magnitude. As a measure of overall sparsity in the predictor $\widehat{\beta}$, we report the ratio $(\sum_j\widehat{\beta}_j^2)^2/\sum_j\widehat{\beta}_j^4$, which serves as a proxy for the effective number of active coefficients. Under the true vector $\beta^\ast$ this ratio takes the value $2.8575$. Lastly, we report average computation times on a standard laptop for all benchmarks. 

\paragraph{Results} 
Values for RMSEs, AUC, the Brier score, and the run times are reported in Appendix~B. \ac{baggls} is compatible with the benchmark methods in terms of out of sample prediction measured by AUC and the Brier score, while more effective at detecting the active terms. Across all simulation scenarios, \ac{baggls} achieves the lowest RMSE on the active (nonzero) entries of $\beta^\ast$ followed by HS. HS imposes stronger shrinkage on zero coefficients as measured by the RMSE on the inactive components, particularly in scenarios with $n=2,000$, where the sample-to-feature ratio $n/p$ is larger than in scenarios with $n=500$. In the application considered in Section~\ref{sec:application} the sample-to-feature ratio $n/p=1.33$ is small.

Table~\ref{tab:sim2} reports the proportion of repetitions in which the three truly active terms are ranked among the top 20 and, separately, among the top 3 estimated effects by absolute magnitude. While L1 and UR do not recover the active effects reliably, HS and \ac{baggls} place all three truly active effects in the top 20 for every run across all scenarios. However, \ac{baggls} outperforms HS when considering only the top 3 terms. In particular, the interaction effect $m_1m_2$ is often missed by HS. for $n=500, d=20$, the intercation appears in the top 3 in 88 of 100 repetitions for HS versus 93 of 100 for \ac{baggls}. Similarly, for $n=500, d=15$, \ac{baggls} detects the interaction 93 times, while HS places it outside the top 3 in 16\% of runs. These results indicate that \ac{baggls} is robust at detecting interaction effects, especially when the sample-to-feature ratio is small. In addition, \ac{baggls} is between 7 and 122 times faster than HS, depending on the scenario.

\begin{table}[tb]
\centering
\begin{tabular}{cccccccc}
& \multicolumn{3}{c}{\% in top 20 terms} & \multicolumn{3}{c}{\% in top 3 terms}& sparsity\\
& $m_1$ & $m_2$ & $m_1m_2$ & $m_1$ & $m_2$ & $m_1m_2$ \\
\hline \hline
\multicolumn{7}{l}{\textit{n=500, d=10, p=56, n/p=8.93}}\\
BaGGLS
& 100\%
& 100\%
& 100\%
& 94\%
& 96\%
& 94\%
& 2.5200 (0.6045)
\\
HS
& 100\%
& 100\%
& 100\%
& 91\%
& 99\%
& 91\%
& 2.2582 (0.5448)
\\
L1
& 81\%
& 91\%
& 81\%
& 53\%
& 75\%
& 53\%
& 6.6936 (2.7878)
\\
UR
& 98\%
& 98\%
& 98\%
& 44\%
& 40\%
& 44\%
& 1.2632 (0.4286)
\\
\hline \hline
\multicolumn{7}{l}{\textit{n=2000, d=10, p=56, n/p=35.71}}\\
BaGGLS
& 100\%
& 100\%
& 100\%
& 100\%
& 100\%
& 100\%
& 2.8241 (0.1598)
\\
HS
& 100\%
& 100\%
& 100\%
& 100\%
& 100\%
& 100\%
& 2.8046 (0.1561)
\\
L1
& 94\%
& 100\%
& 94\%
& 85\%
& 99\%
& 85
& 3.5463 (0.7359)
\\
UR
& 98\%
& 99\%
& 98\%
& 46\%
& 50\%
& 46\%
& 1.2996 (0.5817)
\\
\hline \hline
\multicolumn{7}{l}{\textit{n=500, d=15, p=121, n/p=4.13}}\\
BaGGLS
& 100\%
& 100\%
& 100\%
& 93\%
& 96\%
& 93\%
& 2.3641 (0.6513)
\\
HS
& 100\%
& 100\%
& 100\%
& 84\%
& 98\%
& 84\%
& 2.0703 (0.4973)
\\
L1
& 49\%
& 66\%
& 49\%
& 21\%
& 44\%
& 21\%
& 13.9722 (5.7225)
\\
UR
& 95\%
& 84\%
& 95\%
& 68\%
& 31\%
& 68\%
& 1.6237 (1.0593)
\\
\hline \hline
\multicolumn{7}{l}{\textit{n=2000, d=15, p=121, n/p=16.53}}\\
BaGGLS
& 100\%
& 100\%
& 100\%
& 100\%
& 100\%
& 100\%
& 2.8329 (0.1255)
\\
HS
& 100\%
& 100\%
& 100\%
& 100\%
& 100\%
& 100\%
& 2.8049 (0.1258)
\\
L1
& 88\%
& 99\%
& 88\%
& 61\%
& 92\%
& 61\%
& 5.6560 (1.9977)
\\
UR
& 92\%
& 94\%
& 92\%
& 48\%
& 49\%
& 48\%
& 1.1880 (0.2497)
\\
\hline \hline
\multicolumn{7}{l}{\textit{n=500, d=20, p=211, n/p=2.37}}\\
BaGGLS
& 100\%
& 100\%
& 100\%
& 93\%
& 94\%
& 93\%
& 2.3405 (0.6667)
\\
HS
& 100\%
& 100\%
& 100\%
& 88\%
& 99\%
& 88\%
& 1.9953 (0.4971)
\\
L1
& 13\%
& 19\%
& 13\%
& 8\%
& 13\%
& 8\%
& 31.1999 (10.9530)
\\
UR
& --
& --
& --
& --
& --
& --
& --
\\
\hline \hline
\multicolumn{7}{l}{\textit{n=2000, d=20, p=211, n/p=9.48}}\\
BaGGLS
& 100\%
& 100\%
& 100\%
& 100\%
& 100\%
& 100\%
& 2.8352 (0.1220)
\\
HS
& 100\%
& 100\%
& 100\%
& 100\%
& 100\%
& 100\%
& 2.7953 (0.1115)
\\
L1
& 71\%
& 88\%
& 71\%
& 41\%
& 72\%
& 41\%
& 9.1345 (3.9088)
\\
UR
& 75\%
& 81\%
& 75\%
& 47\%
& 49\%
& 47\%
& 1.3885 (1.0950)
\\
\hline \hline
\end{tabular}
\caption{Simulations. Proportions of runs for which the truly active terms $m_1$, $m_2$, and $m_1m_2$ are among the top 20 or top 3 terms respectively for each benchmark method and all simulation scenarios. The last column reports the average value and the standard deviation (in brackets) for the ratio $(\sum_j\widehat{\beta}_j^2)^2/\sum_j\widehat{\beta}_j^4$. UR did not reliably converge for the scenario $n=500$, $d=20$. \ac{baggls} outperforms the benchmarks in detecting relevant effects including the interaction.}
\label{tab:sim2}
\end{table}

\section{Application to genomic attribution scores}\label{sec:application}
In this section, we illustrate how \ac{baggls} is useful as a post hoc processing method for genomic deep learning explanations. To this end, we apply \ac{baggls} to attribution scores derived from deep learning models trained on synthetic genomic sequences containing real motifs and a known ground truth.

\begin{figure}[tb!]
    \centering\includegraphics[width=0.9\linewidth,keepaspectratio]{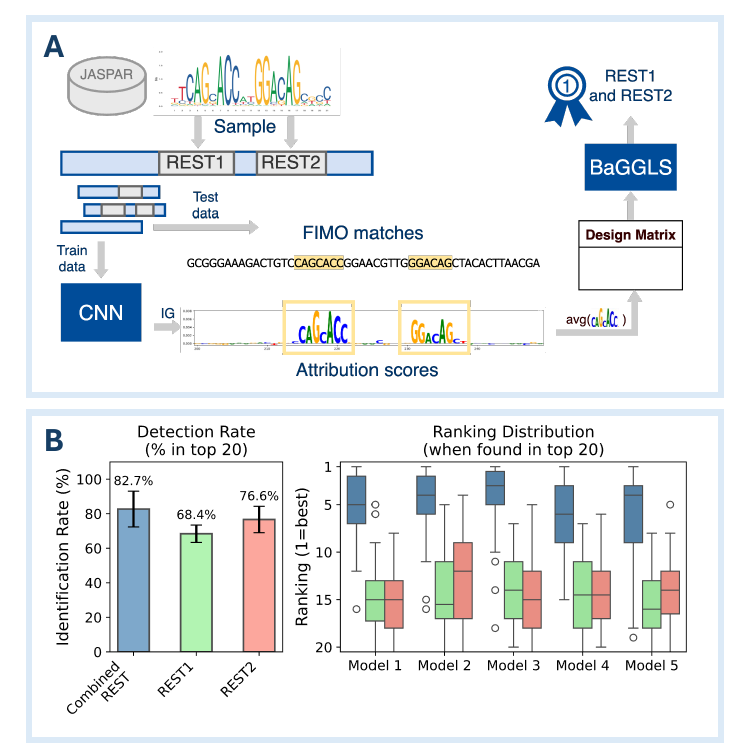}
    \caption{Post-processing of \ac{baggls} on attribution scores. (A) First, we generated synthetic sequences by inserting the sampled composite REST motif in 500bp long random sequences. We trained five shallow CNN models on that data and interpreted the test data with Integrated Gradients. By applying FIMO on the sequences, we obtain matches which resemble motifs from the JASPAR2024 data base. We average the attribution scores in the matched motif region and compute the design matrix and indicator matrix from those scores for \ac{baggls}.(B) We applied \ac{baggls} on the 45 test data sets for each model and measured how often the interaction term was included in the Top 20 terms. \ac{baggls} ranks the interaction term on average in 82.7\% of the test data sets as a top 20 effect while the individual effects in 68.4\% and 76.6\% for the first and second half of the REST motif respectively (left plot). When looking closer to the exact ranking of the identified terms (right plot), the interaction term receives much higher ranking than the individual effects.}
    \label{fig:attr_results}
\end{figure}

\paragraph{Data and motif scanning}
The dataset consists of synthetic DNA sequences with binary labels indicating the presence of a motif set of interest. We follow a similar simulation protocol as described in \cite{Tseng2024} to evaluate the approach on known ground truth. Depending on the defined motif grammar, the motifs are inserted in randomly generated sequence. Here, we explored the REST motif consisting of two submotifs with a specific order and spacing (Figure~\ref{fig:attr_results}A). 35,000 sequences are generated for training and 10,000 for validation, with a sequence length of 500 base pairs. To evaluate \ac{baggls}, we generated additional 45 evaluation datasets with the same grammar each consisting of 2,000 sequences.

FIMO \citep{grant2011fimo} locates motifs by computing significant matches of motifs with position-weight matrices (PWM) from databases in a sequence. Here, we use all latest versions of the human transcription binding site motifs ($d=755$) in JASPAR2024 \citep{rauluseviciute2024jaspar}. Depending on the threshold for the p-value, the results can contain many false positive matches (high threshold) or miss some of the real motif matches (low threshold). We use the default threshold $pthresh=1e^{-4}$ to investigate the robustness of \ac{baggls} by exposing it to noisy motif matches by allowing false positives as well as relevant motifs which are not matched by FIMO despite being present.
\paragraph{Attribution scores from deep learning models}
We train five shallow convolutional neural networks (CNN) on the training data set. We specifically use a simple architecture to not overfit to the synthetic data resulting in an average AUC of 0.91. Details on the architecture and performance can be found in the supplement section \ref{sec:supp_attr}. The trained CNNs are interpreted by the post-hoc attribution method Integrated Gradients \citep{sundararajan2017axiomatic} to obtain position-wise attribution scores which indicate the contribution of one position to the over-all output of an input sequence. We calculate the scores for all evaluation data sets for each of the five CNNs. 
\paragraph{Attribution-based design matrix}
\ac{baggls} requires a pre-defined set of features and interactions passed in a design matrix (Figure~\ref{fig:main}B). Here, we pass all matched motifs by FIMO as possible features to \ac{baggls}. FIMO returns the start and end positions of the matched motifs within each sequence. With that information, we can aggregate the attribution scores in that subregion to obtain motif-wise contributions. Here, we use the absolute average from the attribution scores at the matched motif positions scaled sequence-wise so that the aggregated motif scores add up to 1. To create interaction features, we calculate the pairwise co-activation based on the product of the average contribution scores. We remove rarely occurring features to avoid inefficiency due to a very large number of features by using a cut-off quantile of 0.95 which can be adjusted by the user. The resulting features still comprise of a large number of motif and interaction features (p=1,500), resulting in a ratio of $n\textbackslash p=1.33$ similar to the scenarios considered in Section~\ref{sec:simulations}. The design matrix consists of 755 single effects and 745 interaction effects. Similarly as in the simulation, we fit \ac{baggls} for each of the 45 evaluation data sets for each of the 5 CNNs. In total, this results in 225 different scenarios.

\paragraph{Attribution evaluation}
Due to the sparsity assumption only a small sub-set of regression coefficients $\widehat{\beta}$ derived by \ac{baggls} is meaningfully different from $0$. This is the set of effectively active feature and interaction effects detected by \ac{baggls}. For further interpretation, we calculate the top-20 effects by absolute magnitude of their respective coefficients (see Figure~\ref{fig:main}F). This can be viewed as the set of the most important $20$ effects out of the $p=1,500$ potential effects initially passed to our method. 

For illustration purposes, we consider here data with a known ground truth. This allows us to evaluate whether \ac{baggls} captures the known main interaction, by checking how often the interaction is included within the top-20 effects. \ac{baggls} identifies the interaction term in 82,7\% of the scenarios whereas the REST submotifs were included within the top-20 in 68.4\% and 76.6\% of the scenarios, respectively (Figure~\ref{fig:attr_results}A). It is important to note that when using \ac{baggls} as post-processing method, its performance is highly dependent on the fit of the base CNN. In particular \ac{baggls} cannot detect effects that were overlooked by the base model. When analyzing at the rankings of the terms for each of the five CNN models individually (Figure~\ref{fig:attr_results}B), the interaction term ranks on average higher than the individual submotif effects for each model (Median ranks: interaction term = 4, REST1 = 15, REST2 = 14). This shows that \ac{baggls} robustly identifies the driving interaction among noisy attribution scores and prioritizes the interaction over individual effects. For this reason, \ac{baggls} can be used as highly interpretable post-hoc processing method for trained deep learning genomic classifiers. Identifying important interactions is not directly possible from the black-box CNNs, but crucial for understanding the underlying biological process. 

\section{Discussion}\label{sec:discussion}

We propose \ac{baggls}, an interaction detection method for high-dimensional and noisy biological data. To this end, we make the following main contributions. (i) We propose to use a structured probit regression model as a post-hoc analysis tool for deep classifiers. Our model takes a large number of potential effect, including interactions, as input and identifies a small subset of important and truly active effects. (ii) Sparsity is imposed through a novel shrinkage prior that respects the overlapping group structure resulting from the inclusion of many interaction terms in the additive predictor. (iii) We propose a fast algorithm for posterior inference based on a unified skewed-normal approximation. This is crucial for scalability to scenarios with many potential effects and a a small sample-to-feature ratio. (iv) We conduct an extensive simulation study, and show that \ac{baggls} outperforms state-of-the-art benchmarks in detecting relevant interaction effects. (v) We illustrate the merits of our approach in an application on motif interactions in genomic data which is known to be complex and noisy offering a good use case for comparisons. 

Even though, our general framework allows for arbitrary and more complex effect types to be specified, we have only considered continuous features and interactions based on co-occurrences within our application. The inclusion of more complex effects, for example, by including binary features and boolean interactions in the context of logic regression \citep{RucKooLeb2003} coupled with regulatory logic in genomics \citep{buchler2003schemes} is a promising direction. 

Combining our novel overlapping group shrinkage prior with regression models beyond the probit model, is a further avenue for future research. For example, current deep learning prediction tasks in the genomic domain shift from classification to continuous and categorical responses by considering biological signals like gene expression or read counts directly. Quantitative readouts provide a more fine-grained picture of a biological signal. Extending \ac{baggls} to continuous responses might be one pathway to account for low-affinity motifs which are sometimes ignored due to thresholds by peak calling methods \citep{zeitlinger2020seven}.

While we demonstrate application to motif detection, other biological domains can also benefit from \ac{baggls}. In genetics, single-nucleotide polymorphisms are studied in the context of disease. While rare disorders are frequently driven by individual variants with large effects, common-variant contributions to complex traits are highly polygenic and often involve context-dependent effects that are hard to resolve in high-dimensional genotype space \citep{wray2018common}. Currently, regulatory genomics is shifting from bulk assays to single-cell and related high-resolution assays. These provide finer cell-type specificity but produce sparser, noisier data matrices that change the statistical challenges and opportunities \citep{bouland2023consequences}. This shift opens new applications for \ac{baggls}.

However, there are also limitations visible in the presented use case. In some genomic regions, composite regulatory elements, which are arrangements of multiple nearby binding sites, can alter local sequence preferences and functional readout \citep{jolma2015dna}. Bulk models that assume independent binding sites often miss these dependencies, which leads to only a sparse but important set of interactions being captured. Adjusting the design matrix to include interaction terms based on experimentally validated composite elements can therefore make interaction interpretations more precise. This can further be improved by an iterative approach in which interpretable features from deep models, for example, convolutional filters \citep{Tseng2024} or in-silico perturbation readouts \citep{gjoni2024supremo}, are used to detect candidate composite arrangements that are then incorporated into the design matrix to refine sparsity structure and interaction estimates. The interaction patterns detected by our method could also serve as hypotheses for previously unrecognized composite elements or motif interactions. These hypotheses can be ranked by \ac{baggls} and subsequently tested in targeted biochemical or reporter assays, providing a systematic way to identify novel regulatory interactions.
 
\FloatBarrier
\setlength{\bibsep}{0pt plus 0.3ex}
\bibliography{litlist}

@article{XuSchMakQiaHop2017,
  title={{B}ayesian sparse global-local shrinkage regression for selection of grouped variables},
  author={Xu, Zemei and Schmidt, Daniel F and Makalic, Enes and Qian, Guoqi and Hopper, John L},
  journal={arXiv preprint arXiv:1709.04333},
  year={2017}
}

@article{RucKooLeb2003,
  title={Logic regression},
  author={Ruczinski, Ingo and Kooperberg, Charles and LeBlanc, Michael},
  journal={Journal of Computational and graphical Statistics},
  volume={12},
  number={3},
  pages={475--511},
  year={2003},
  publisher={Taylor \& Francis}
}

@article{FasDurZan2022,
  title={Scalable and accurate variational {B}ayes for high-dimensional binary regression models},
  author={Fasano, Augusto and Durante, Daniele and Zanella, Giacomo},
  journal={Biometrika},
  volume={109},
  number={4},
  pages={901--919},
  year={2022},
  publisher={Oxford University Press}
}

@article{AreAzz2006,
author = {Arrelano-Valle, R. B. and Azzalini, A.},
title = {On the Unification of Families of Skew-normal Distributions},
journal = {Scandinavian Journal of Statistics},
volume = {33},
number = {3},
pages = {561-574},
year = {2006}
}

@article{DurRig2019,
author = {Daniele Durante and Tommaso Rigon},
title = {{Conditionally Conjugate Mean-Field Variational {B}ayes for Logistic Models}},
volume = {34},
journal = {Statistical Science},
number = {3},
publisher = {Institute of Mathematical Statistics},
pages = {472 -- 485},
year = {2019},
doi = {10.1214/19-STS712},
}

@article{PozDurSza2024,
  title={Skew-symmetric approximations of posterior distributions},
  author={Pozza, Francesco and Durante, Daniele and Szabo, Botond},
  journal={Journal of the Royal Statistical Society, Series B: Statistical Methodology},
  year={2025},
note = {in press}
}

@article{BhaChaMal2016,
    author = {Bhattacharya, Anirban and Chakraborty, Antik and Mallick, Bani K.},
    title = {Fast sampling with {G}aussian scale mixture priors in high-dimensional regression},
    journal = {Biometrika},
    volume = {103},
    number = {4},
    pages = {985-991},
    year = {2016},
    month = {10},
    issn = {0006-3444},
}

@article{Bue2017,
  title={brms: An {R} package for {B}ayesian multilevel models using {Stan}},
  author={B{\"u}rkner, Paul-Christian},
  journal={Journal of statistical software},
  volume={80},
  pages={1--28},
  year={2017}
}

@article{IshRao2005,
author = {Ishwaran, Hemant and Rao, J. Sunil},
title = {{Spike and slab variable selection: Frequentist and {B}ayesian strategies}},
volume = {33},
journal = {The Annals of Statistics},
number = {2},
publisher = {Institute of Mathematical Statistics},
pages = {730 -- 773},
year = {2005}
}

@incollection{PolSco2011,
    author = {Polson, Nicholas G. and Scott, James G.},
    isbn = {9780199694587},
    title = {Shrink Globally, Act Locally: Sparse {B}ayesian Regularization and Prediction},
    booktitle = {{B}ayesian Statistics 9},
    publisher = {Oxford University Press},
    year = {2011},
    month = {10}
}

@article{ParCas2008,
  title={The {B}ayesian lasso},
  author={Park, Trevor and Casella, George},
  journal={Journal of the american statistical association},
  volume={103},
  number={482},
  pages={681--686},
  year={2008},
  publisher={Taylor \& Francis}
}

@article{TanChe2025,
  title={Variational inference based on a subclass of closed skew normals},
  author={Tan, Linda SL and Chen, Aoxiang},
  journal={Journal of Computational and Graphical Statistics},
  volume={34},
  number={2},
  pages={422--436},
  year={2025},
  publisher={Taylor \& Francis}
}

@article{RaySza2022,
  title={Variational {B}ayes for high-dimensional linear regression with sparse priors},
  author={Ray, Kolyan and Szab{\'o}, Botond},
  journal={Journal of the American Statistical Association},
  volume={117},
  number={539},
  pages={1270--1281},
  year={2022},
  publisher={Taylor \& Francis}
}

@book{Bis2006,
  title={Pattern recognition and machine learning},
  author={Bishop, Christopher M},
  year={2006},
  publisher={Springer}
}

@book{giraud2021introduction,
    author = {Giraud, C.},
    title = {Introduction to High-Dimensional Statistics},
    publisher = {Chapman and Hall/CRC},
    year = {2021}
}

@article{Dur2019,
  title={Conjugate {B}ayes for probit regression via unified skew-normal distributions},
  author={Durante, Daniele},
  journal={Biometrika},
  volume={106},
  number={4},
  pages={765--779},
  year={2019},
  publisher={Oxford University Press}
}

@article{ZhaXuZha2019,
  title={A novel variational {B}ayesian method for variable selection in logistic regression models},
  author={Zhang, Chun-Xia and Xu, Shuang and Zhang, Jiang-She},
  journal={Computational Statistics \& Data Analysis},
  volume={133},
  pages={1--19},
  year={2019},
  publisher={Elsevier}
}

@inproceedings{RaySzaCla2020,
 title = {Spike and slab variational {B}ayes for high dimensional logistic regression},
 author = {Ray, Kolyan and Szabo, Botond and Clara, Gabriel},
 booktitle = {Advances in Neural Information Processing Systems},
 editor = {H. Larochelle and M. Ranzato and R. Hadsell and M.F. Balcan and H. Lin},
 pages = {14423--14434},
 publisher = {Curran Associates, Inc.},
 volume = {33},
 year = {2020}
}

@article{AncFasDurZan2023,
  title={{B}ayesian conjugacy in probit, tobit, multinomial probit and extensions: A review and new results},
  author={Anceschi, Niccol{\`o} and Fasano, Augusto and Durante, Daniele and Zanella, Giacomo},
  journal={Journal of the American Statistical Association},
  volume={118},
  number={542},
  pages={1451--1469},
  year={2023},
  publisher={Taylor \& Francis}
}

@article{KocTanBanNot2025,
  title={Variational inference for hierarchical models with conditional scale and skewness corrections},
  author={Kock, Lucas and Tan, Linda SL and Bansal, Prateek and Nott, David J},
  journal={arXiv preprint arXiv:2503.18075},
  year={2025}
}

@article{DurPozSza2024,
  title={Skewed {B}ernstein--von {M}ises theorem and skew-modal approximations},
  author={Durante, Daniele and Pozza, Francesco and Szabo, Botond},
  journal={The Annals of Statistics},
  volume={52},
  number={6},
  pages={2714--2737},
  year={2024},
  publisher={Institute of Mathematical Statistics}
}

@article{GneRaf2007,
  title={Strictly proper scoring rules, prediction, and estimation},
  author={Gneiting, Tilmann and Raftery, Adrian E},
  journal={Journal of the American statistical Association},
  volume={102},
  number={477},
  pages={359--378},
  year={2007},
  publisher={Taylor \& Francis}
}

@inproceedings{SeaPer2010,
  title={statsmodels: Econometric and statistical modeling with python},
  author={Seabold, Skipper and Perktold, Josef},
  booktitle={9th Python in Science Conference},
  year={2010},
}

@article{CarPolSco2010,
    author = {Carvalho, Carlos M. and Polson, Nicholas G. and Scott, James G.},
    title = {The horseshoe estimator for sparse signals},
    journal = {Biometrika},
    volume = {97},
    number = {2},
    pages = {465-480},
    year = {2010},
    month = {04},
    issn = {0006-3444},
}

@inproceedings{Tseng2024,
  title={A mechanistically interpretable neural-network architecture for discovery of regulatory genomics},
  author={Tseng, Alex M and Eraslan, G{\"o}kcen and Diamant, Nathaniel Lee and Biancalani, Tommaso and Scalia, Gabriele},
  booktitle={ICLR 2024 Workshop on Machine Learning for Genomics Explorations},
  year={2024}
}

@article{shrikumar2018technical,
  title={Technical note on transcription factor motif discovery from importance scores (TF-MoDISco) version 0.5. 6.5},
  author={Shrikumar, Avanti and Tian, Katherine and Avsec, {\v{Z}}iga and Shcherbina, Anna and Banerjee, Abhimanyu and Sharmin, Mahfuza and Nair, Surag and Kundaje, Anshul},
  journal={arXiv preprint arXiv:1811.00416},
  year={2018}
}

@article{grant2011fimo,
  title={FIMO: scanning for occurrences of a given motif},
  author={Grant, Charles E and Bailey, Timothy L and Noble, William Stafford},
  journal={Bioinformatics},
  volume={27},
  number={7},
  pages={1017--1018},
  year={2011},
  publisher={Oxford University Press}
}

@article{novakovsky2023obtaining,
  title={Obtaining genetics insights from deep learning via explainable artificial intelligence},
  author={Novakovsky, Gherman and Dexter, Nick and Libbrecht, Maxwell W and Wasserman, Wyeth W and Mostafavi, Sara},
  journal={Nature Reviews Genetics},
  volume={24},
  number={2},
  pages={125--137},
  year={2023},
  publisher={Nature Publishing Group UK London}
}

@article{BleKucMca2017,
  title={Variational inference: A review for statisticians},
  author={Blei, David M and Kucukelbir, Alp and McAuliffe, Jon D},
  journal={Journal of the American statistical Association},
  volume={112},
  number={518},
  pages={859--877},
  year={2017},
  publisher={Taylor \& Francis}
}

@article{YanBonRei2025,
author = {Yanchenko, Eric and Bondell, Howard D.and Reich, Brian J.},
title = {The {R2D2} Prior for Generalized Linear Mixed Models},
journal = {The American Statistician},
volume = {79},
number = {1},
pages = {40--49},
year = {2025},
publisher = {ASA Website}
}

@article{GeoMcc1997,
  title={Approaches for {B}ayesian variable selection},
  author={George, Edward I and McCulloch, Robert E},
  journal={Statistica sinica},
  pages={339--373},
  year={1997},
  publisher={JSTOR}
}

@article{LiaPauMolClyBer2008,
  title={Mixtures of g priors for {B}ayesian variable selection},
  author={Liang, Feng and Paulo, Rui and Molina, German and Clyde, Merlise A and Berger, Jim O},
  journal={Journal of the American Statistical Association},
  volume={103},
  number={481},
  pages={410--423},
  year={2008},
  publisher={Taylor \& Francis}
}

@article{borah2024review,
  title={A review on advancements in feature selection and feature extraction for high-dimensional NGS data analysis},
  author={Borah, Kasmika and Das, Himanish Shekhar and Seth, Soumita and Mallick, Koushik and Rahaman, Zubair and Mallik, Saurav},
  journal={Functional \& Integrative Genomics},
  volume={24},
  number={5},
  pages={139},
  year={2024},
  publisher={Springer}
}

@article{yang2021feature,
  title={Feature selection revisited in the single-cell era},
  author={Yang, Pengyi and Huang, Hao and Liu, Chunlei},
  journal={Genome Biology},
  volume={22},
  number={1},
  pages={321},
  year={2021},
  publisher={Springer}
}

@article{rauluseviciute2024jaspar,
  title={JASPAR 2024: 20th anniversary of the open-access database of transcription factor binding profiles},
  author={Rauluseviciute, Ieva and Riudavets-Puig, Rafael and Blanc-Mathieu, Romain and Castro-Mondragon, Jaime A and Ferenc, Katalin and Kumar, Vipin and Lemma, Roza Berhanu and Lucas, J{\'e}r{\'e}my and Ch{\`e}neby, Jeanne and Baranasic, Damir and others},
  journal={Nucleic acids research},
  volume={52},
  number={D1},
  pages={D174--D182},
  year={2024},
  publisher={Oxford University Press}
}

@article{van2024designing,
  title={Designing interpretable deep learning applications for functional genomics: a quantitative analysis},
  author={Van Hilten, Arno and Katz, Sonja and Saccenti, Edoardo and Niessen, Wiro J and Roshchupkin, Gennady V},
  journal={Briefings in Bioinformatics},
  volume={25},
  number={5},
  year={2024},
  publisher={Oxford Academic}
}

@article{ismail2025deep,
  title={Deep Learning for Regulatory Genomics: A Survey of Models, Challenges, and Applications},
  author={Ismail, Fathima Nuzla and Sengupta, Abira and Amarasoma, Shanika},
  journal={Bioinformatics Advances},
  pages={vbaf271},
  year={2025},
  publisher={Oxford University Press}
}

@article{xie2025dna,
  title={DNA-guided transcription factor interactions extend human gene regulatory code},
  author={Xie, Zhiyuan and Sokolov, Ilya and Osmala, Maria and Yue, Xue and Bower, Grace and Pett, J Patrick and Chen, Yinan and Wang, Kai and Cavga, Ayse Derya and Popov, Alexander and others},
  journal={Nature},
  pages={1--10},
  year={2025},
  publisher={Nature Publishing Group UK London}
}

@article{zeitlinger2020seven,
  title={Seven myths of how transcription factors read the cis-regulatory code},
  author={Zeitlinger, Julia},
  journal={Current opinion in systems biology},
  volume={23},
  pages={22--31},
  year={2020},
  publisher={Elsevier}
}

@article{bartoszewicz2021,
    author = {Bartoszewicz, Jakub M and Seidel, Anja and Renard, Bernhard Y},
    title = {Interpretable detection of novel human viruses from genome sequencing data},
    journal = {NAR Genomics and Bioinformatics},
    volume = {3},
    number = {1},
    pages = {lqab004},
    year = {2021},
    month = {02},
    doi = {10.1093/nargab/lqab004},
    eprint = {https://academic.oup.com/nargab/article-pdf/3/1/lqab004/36165658/lqab004.pdf},
}

@article{forsberg2017accounting,
  title={Accounting for genetic interactions improves modeling of individual quantitative trait phenotypes in yeast},
  author={Forsberg, Simon KG and Bloom, Joshua S and Sadhu, Meru J and Kruglyak, Leonid and Carlborg, {\"O}rjan},
  journal={Nature genetics},
  volume={49},
  number={4},
  pages={497--503},
  year={2017},
  publisher={Nature Publishing Group US New York}
}

@article{wheeler2016survey,
  title={Survey of the heritability and sparse architecture of gene expression traits across human tissues},
  author={Wheeler, Heather E and Shah, Kaanan P and Brenner, Jonathon and Garcia, Tzintzuni and Aquino-Michaels, Keston and GTEx Consortium and Cox, Nancy J and Nicolae, Dan L and Im, Hae Kyung},
  journal={PLoS genetics},
  volume={12},
  number={11},
  pages={e1006423},
  year={2016},
  publisher={Public Library of Science San Francisco, CA USA}
}

@article{bouland2023consequences,
  title={Consequences and opportunities arising due to sparser single-cell RNA-seq datasets},
  author={Bouland, Gerard A and Mahfouz, Ahmed and Reinders, Marcel JT},
  journal={Genome biology},
  volume={24},
  number={1},
  pages={86},
  year={2023},
  publisher={Springer}
}

@article{wray2018common,
  title={Common disease is more complex than implied by the core gene omnigenic model},
  author={Wray, Naomi R and Wijmenga, Cisca and Sullivan, Patrick F and Yang, Jian and Visscher, Peter M},
  journal={Cell},
  volume={173},
  number={7},
  pages={1573--1580},
  year={2018},
  publisher={Elsevier}
}

@article{jolma2015dna,
  title={DNA-dependent formation of transcription factor pairs alters their binding specificity},
  author={Jolma, Arttu and Yin, Yimeng and Nitta, Kazuhiro R and Dave, Kashyap and Popov, Alexander and Taipale, Minna and Enge, Martin and Kivioja, Teemu and Morgunova, Ekaterina and Taipale, Jussi},
  journal={Nature},
  volume={527},
  number={7578},
  pages={384--388},
  year={2015},
  publisher={Nature Publishing Group UK London}
}

@article{gjoni2024supremo,
  title={SuPreMo: a computational tool for streamlining in silico perturbation using sequence-based predictive models},
  author={Gjoni, Ketrin and Pollard, Katherine S},
  journal={Bioinformatics},
  volume={40},
  number={6},
  pages={btae340},
  year={2024},
  publisher={Oxford University Press}
}

@inproceedings{sundararajan2017axiomatic,
  title={Axiomatic attribution for deep networks},
  author={Sundararajan, Mukund and Taly, Ankur and Yan, Qiqi},
  booktitle={International conference on machine learning},
  pages={3319--3328},
  year={2017},
  organization={PMLR}
}

@article{buchler2003schemes,
  title={On schemes of combinatorial transcription logic},
  author={Buchler, Nicolas E and Gerland, Ulrich and Hwa, Terence},
  journal={Proceedings of the National Academy of Sciences},
  volume={100},
  number={9},
  pages={5136--5141},
  year={2003},
  publisher={National Academy of Sciences}
}

@article{majdandzic2023correcting,
  title={Correcting gradient-based interpretations of deep neural networks for genomics},
  author={Majdandzic, Antonio and Rajesh, Chandana and Koo, Peter K},
  journal={Genome Biology},
  volume={24},
  number={1},
  pages={109},
  year={2023},
  publisher={Springer}
}
\FloatBarrier
\appendix
\section{Variational Inference}

The full vector of the $n+3p+2d+2$ unknown model parameters is $\theta=\left(z^\top,\beta^\top,\tau,\nu,\lambda^\top,c^\top,\delta^\top,t^\top\right)$. We consider the following variational approximation
\begin{align}\label{eq: q}
q(\theta)=q(\beta\mid z)\left(\prod_{i=1}^nq(z_i)\right)q(\tau)q(\nu)\left(\prod_{j=1}^p q(\lambda_j)q(c_j)\right)\left(\prod_{l=1}^d q(\delta_l)q(t_l)\right).
\end{align}
Note that \eqref{eq: q} is not the fully factorized mean-field approximation as we do neither assume independence between $\beta$ and $z$ nor between the individual entries of $\beta$. Instead we consider the partially factorized approximation introduced by \citet{FasDurZan2022}. The structure \eqref{eq: q} implies that $q^\ast(\beta\mid z)$ is a $p$-dimensional Gaussian distribution, $q^\ast(z_i)$ is a truncated normal distribution and all remaining factors of $q^\ast(\theta)$ are inverse gamma distributions. We can thus write

\begin{align*}
    q^\ast(\beta\mid z) &= \phi_p(\beta;B(\beta)z,\Sigma(\beta));\\
    q^\ast(z_i)&=1_{(2y_i-1)z_i>0}\frac{\phi(z_i;\mu(z_i),\sigma^2(z_i)}{\Phi((2y_i-1)\mu(z_i)(\sigma^2(z_i))^{-\frac{1}{2}})} &\qquad&i=1,\dots,n;\\
    q^\ast(\tau)&=p_{\IGD}(\tau;a(\tau),b(\tau));\\
    q^\ast(\nu)&=p_{\IGD}(\nu;a(\nu),b(\nu));\\
    q^\ast(\lambda_j)&=p_{\IGD}(\lambda_j;a(\lambda_j),b(\lambda_j))&\qquad&j=1,\dots,p;\\
    q^\ast(c_j)&=p_{\IGD}(c_j;a(c_j),b(c_j))&\qquad&j=1,\dots,p;\\
    q^\ast(\delta_l)&=p_{\IGD}(\delta_l;a(\delta_l),b(\delta_l))&\qquad&l=1,\dots,d;\\
    q^\ast(t_l)&=p_{\IGD}(t_l;a(t_l),b(t_l))&\qquad&l=1,\dots,d.
\end{align*}

The optimal variational parameters are given as
\begin{align*}
    \Sigma(\beta)&=\left(X^\top X+\diag\left(\frac{a(\tau)a(\lambda_1)}{b(\tau)b(\lambda_1)}\prod_{J_{1l}=1}\frac{a(\delta_l)}{b(\delta_l)},\dots,\frac{a(\tau)a(\lambda_p)}{b(\tau)b(\lambda_p)}\prod_{J_{pl}=1}\frac{a(\delta_l)}{b(\delta_l)}\right)\right)^{-1}\\
    B(\beta)&=\Sigma(\beta)X^\top\\
    \mu(z_i)&=\sigma^2(z_i)x_i^\top\Sigma(\beta)X_{-i}^\top(\mathbb{E}_q[z_1],\dots,\mathbb{E}_q[z_{i-1}],\mathbb{E}_q[z_{i+1}],\dots,\mathbb{E}_q[z_n])^\top\\
    \sigma^2(z_i)&=\left(1-x_i^\top\Sigma(\beta)x_i\right)^{-1}\\
    a(\tau)&=\frac{p+1}{2}\\
    b(\tau)&=\frac{1}{2}\sum_{j=1}^p\left(\mathbb{E}_q[\beta_j^2]\frac{a(\lambda_j)}{b(\lambda_j)}\prod_{J_{jl}=1}\frac{a(\delta_l)}{b(\delta_l)}\right)+\frac{a(\nu)}{b(\nu)}\\
    a(\nu)&=1\\
    b(\nu)&=\frac{a(\tau)}{b(\tau)}+1\\
    a(\lambda_j)&=1\\
    b(\lambda_j)&=\frac{1}{2}\mathbb{E}_q[\beta_j^2]\frac{a(\tau)}{b(\tau)}\prod_{\mathbf{J}_{jl}=1}\frac{a(\delta_l)}{b(\delta_l)}+\frac{a(c_j)}{b(c_j)}\\
    a(c_j)&=1\\
    b(c_j)&=\frac{a(\lambda_j)}{b(\lambda_j)}+1\\
    a(\delta_l)&=\frac{\sum_{j=1}^pJ_{jl}+1}{2}\\
    b(\delta_l)&=\frac{a(\tau)}{b(\tau)}\sum_{j=1}^pJ_{jl}\frac{a(\lambda_l)}{b(\lambda_l)}\mathbb{E}_q[\beta_j^2]+\frac{a(t_l)}{b(t_l)}\\
    a(t_l)&=1\\
    b(t_l)&=1+\frac{a(\delta_l)}{b(\delta_l)},
\end{align*}
where 
\begin{align*}
\mathbf{E}_q[\beta_j^2]=&\left(\Sigma(\beta)+\Sigma(\beta)X^\top\diag\left(\sigma^2(z_1)-(\mathbf{E}[z_1]-\mu(z_1))\mathbf{E}[z_1],\dots,\sigma^2(z_n)-(\mathbf{E}[z_n]-\mu(z_n))\mathbf{E}[z_n]\right)X\Sigma(\beta)\right)_{jj}\\&+\left(B(\beta)\left(\mathbf{E}[z_1],\dots,\mathbf{E}[z_n]\right)^\top\right)_{j}^2;\\
    \mathbf{E}_q[z_i]=&\mu(z_i)+(2y_i-1)\sqrt{\sigma^2(z_i)}\frac{\phi\left(\mu(z_i)\sigma^2(z_i)^{-\frac{1}{2}}\right)}{\Phi\left((2y_i-1)\mu(z_i)\sigma^2(z_i)^{-\frac{1}{2}}\right)}.
\end{align*}
We refer to \citet{FasDurZan2022} for technical details in the derivation of $q^\ast(\beta,z)$. Using this set of equations $q^\ast(\theta)$ can be updated via coordinate ascent \citep{Bis2006}. Note that each step involves inverting the $p\times p$ dimensional matrix $\Sigma(\beta)$. If $p>n$, the Woodbury matrix identity is used to invert a lower dimensional $n\times n$ matrix instead. 

\section{Additional simulation results}\label{app:sim}
Table~\ref{tab:sim1} summarizes the overall RMSE, the RMSE computed separately on the active (nonzero) and inactive (zero) entries of $\beta^\ast$, the area under the receiver operating characteristic curve (AUC), and the Brier score, $n^{-1}\sum_{i=1}^n(\Phi(x_i^\top\widehat{\beta})-y_i)^2$, evaluated on an additional hold-out test data set with $n=10,000$ observations. We report average values as well as standard deviations (in brackets) over 100 repetitions. The table also reports the average run-time on a standard laptop  for each of the six simulation scenarios considered. 

\begin{table}[tb]
\centering
\begin{adjustbox}{max width=\textwidth}
\begin{tabular}{ccccccc}
& RMSE ($\downarrow$) & RMSE non zero ($\downarrow$)& RMSE true-zero ($\downarrow$) & AUC ($\uparrow$)& Brier score ($\downarrow$)&  time (in s)\\
\hline \hline
\multicolumn{7}{l}{\textit{n=500, d=10, p=56, n/p=8.93}}\\
BaGGLS
& \textbf{0.7022} (0.4180)
& \textbf{0.5627} (0.3889)
& \textbf{0.3881} (0.2220)
& 0.8648 (0.0040)
& 0.1483 (0.0026)
& 0.5866 \\
HS
& 0.8052 (0.3671)
& 0.6876 (0.3388)
& \textbf{0.3964} (0.1960)
& \textbf{0.8660} (0.0031)
& \textbf{0.1478} (0.0020)
& 71.5908 \\
L1
& 1.3827 (0.2706)
& 0.6892 (0.2880)
& 1.1727 (0.2278)
& 0.8457 (0.0063)
& 0.1616 (0.0042)
& 0.3983 \\
UR
& 114.2356 (85.8997)
& 14.4545 (10.4597)
& 113.2603 (85.3364)
& 0.7569 (0.0845)
& 0.2601 (0.0839)
& 0.0238 \\
\hline \hline
\multicolumn{7}{l}{\textit{n=2000, d=10, p=56, n/p=35.71}}\\
BaGGLS
& \textbf{0.2356} (0.1305)
& \textbf{0.1775} (0.1159)
& \textbf{0.1381} (0.0924)
& 0.8674 (0.0013)
& 0.1472 (0.0007)
& 10.2991 \\
HS
& \textbf{0.2186} (0.1247)
& \textbf{0.1737} (0.1196)
& \textbf{0.1191} (0.0683)
& \textbf{0.8676} (0.0010)
& \textbf{0.1471} (0.0006)
& 250.9252 \\
L1
& 0.7699 (0.1804)
& 0.4266 (0.2162)
& 0.6206 (0.1074)
& 0.8611 (0.0019)
& 0.1510 (0.0012)
& 0.4775\\
UR
& 59.8707 (44.3895)
& 7.3706 (5.4846)
& 59.4063 (44.0615)
& 0.8398 (0.0245)
& 0.1736 (0.0245)
& 4.3269 \\
\hline \hline
\multicolumn{7}{l}{\textit{n=500, d=15, p=121, n/p=4.13}}\\
BaGGLS
& \textbf{0.8325} (0.4839)
& \textbf{0.6967} (0.4367)
& \textbf{0.4219} (0.2704)
& 0.8533 (0.0058)
& \textbf{0.1558} (0.0036)
& 7.8429 \\
HS
& \textbf{0.9311} (0.3800)
& 0.8220 (0.3459)
& \textbf{0.4177} (0.2040)
& \textbf{0.8558} (0.0035)
& \textbf{0.1545} (0.0023)
& 90.8542 \\
L1
& 2.3274 (0.3461)
& 0.8842 (0.3459)
& 2.1265 (0.3364)
& 0.8056 (0.0100)
& 0.1956 (0.0073)
& 6.7732 \\
UR
& 86.4223 (67.9976)
& 11.6303 (7.5754)
& 85.0680 (68.2881)
& 0.7597 (0.0383)
& 0.2590 (0.0384)
& 5.9180 \\
\hline \hline
\multicolumn{7}{l}{\textit{n=2000, d=15, p=121, n/p=16.53}}\\
BaGGLS
& 0.2267 (0.1152)
& \textbf{0.1673} (0.1031)
& 0.1410 (0.0787)
& 0.8614 (0.0012)
& 0.1504 (0.0007)
& 26.7289 \\
HS
& \textbf{0.1994} (0.1115)
& \textbf{0.1653} (0.1063)
& \textbf{0.1029} (0.0546)
& \textbf{0.8621} (0.0007)
& \textbf{0.1500} (0.0004)
& 298.2052 \\
L1
& 1.2030 (0.2149)
& 0.5456 (0.2493)
& 1.0519 (0.1639)
& 0.8459 (0.0029)
& 0.1603 (0.0018)
& 5.7420 \\
UR
& 53.0565 (33.0476)
& 4.9558 (2.7377)
& 52.8008 (32.9721)
& 0.8287 (0.0162)
& 0.1788 (0.0174)
& 6.8981 \\
\hline \hline
\multicolumn{7}{l}{\textit{n=500, d=20, p=211, n/p=2.37}}\\
BaGGLS
& \textbf{0.8575} (0.4544)
& \textbf{0.7135} (0.4310)
& 0.4288 (0.2511)
& 0.8476 (0.0076)
& 0.1593 (0.0044)
& 15.6204 \\
HS
& \textbf{0.9166} (0.3401)
& 0.8283 (0.3253)
& \textbf{0.3707} (0.1627)
& \textbf{0.8516} (0.0035)
& \textbf{0.1571} (0.0021)
& 107.5822 \\
L1
& 3.9889 (0.5015)
& 1.3117 (0.3719)
& 3.7540 (0.4599)
& 0.7648 (0.0114)
& 0.2447 (0.0091)
& 71.3119 \\
UR
& --
& --
& --
& --
& --
& -- \\
\hline \hline
\multicolumn{7}{l}{\textit{n=2000, d=20, p=211, n/p=9.48}}\\
BaGGLS
& 0.2388 (0.1003)
& \textbf{0.1709} (0.0942)
& 0.1528 (0.0754)
& 0.8565 (0.0016)
& 0.1536 (0.0009)
& 48.2972 \\
HS
& \textbf{0.2050} (0.0984)
& \textbf{0.1710} (0.0945)
& \textbf{0.1045} (0.0513)
& \textbf{0.8576} (0.0007)
& \textbf{0.1529} (0.0005)
& 371.5898 \\
L1
& 1.7003 (0.2597)
& 0.7289 (0.2766)
& 1.5154 (0.2332)
& 0.8294 (0.0038)
& 0.1724 (0.0027)
& 34.3788 \\
UR
& 37.2856 (24.8385)
& 3.4586 (2.0772)
& 37.0293 (24.8942)
& 0.8184 (0.0096)
& 0.1826 (0.0080)
& 12.4084 \\
\hline \hline
\end{tabular}
\end{adjustbox}
\caption{Simulations. Average values and standard deviations (in brackets) across all independent repetitions for the six simulation scenarios for the overall RMSE, the RMSE computed on the active (nonzero) and inactive (zero) entries of $\beta^\ast$, the area under the receiver operating characteristic curve (AUC), and the Brier score, as well as the average computation times. Bold values indicate the best performing method in terms of the best mean value as well as methods potentially tied according to a one-sided t-test with level $\alpha=0.05$. UR did not reliably converge for the scenario $n=500, d=20$.}
\label{tab:sim1}
\end{table}

\section{Application to genomic attribution scores}
\label{sec:supp_attr}
This section gives additional details on the deep learning models considered within our application presented in Section~4 of the main paper.
The architecture first employs a one-dimensional convolutional layer with 8 filters (kernel size = 8), designed to scan the input sequences for motif patterns. The output of this layer is processed by a ReLU activation function and batch normalization. This is followed by two fully connected layers, each with 16 units and ReLU activations. To mitigate overfitting, a dropout layer with a rate of 0.3 is applied after each fully connected layer. The architecture concludes with a dense output layer followed by a sigmoid activation function, which outputs a probability score representing the likelihood of the input sequence containing the target motif set. We trained 5 models and evaluated the predictive performance on a separate test data set consisting of 2000 samples (see Table \ref{tab:cnn_perform}).\\
\begin{table}[h]
\begin{tabular}{|l|l|l|l|l|l|}
\hline
Model: & Model 1 & Model 2 & Model 3 & Model 4 & Model 5 \\ \hline
AUC  & 0.9157 ± 0.0074 & 0.9137 ± 0.0060  &  0.9037 ± 0.0069 & 0.9088 ± 0.0062  &  0.9099 ± 0.0067  \\ \hline
\end{tabular}
\caption{Performance of the CNNs on the 45 evaluation sequence datasets with each 2000 sequences. AUC are calculated on each dataset separately, since they were used separately for fitting BaGGLS. On average, an AUC of 0.91 could be achieved.}
\label{tab:cnn_perform}
\end{table}

\end{document}